\documentclass[runningheads]{llncs}
\usepackage{graphicx}
\begin{document}
\title{A Sequential Framework for Detection and Classification of Abnormal Teeth in Panoramic X-rays}
%
%
\author{Tudor Dascalu\inst{1}\orcidID{0000-0002-4942-1389} \and Shaqayeq Ramezanzade\inst{2}\orcidID{0000-0002-1012-0730} \and Azam Bakhshandeh\inst{2}\orcidID{0000-0003-2974-1331} \and Lars Bjørndal\inst{2}\orcidID{0000-0002-2183-6400} \and
Bulat Ibragimov\inst{1}\orcidID{0000-0001-7739-7788}}
\authorrunning{T. Dascalu and B. Ibragimov}
%
\institute{Department of Computer Science, University of Copenhagen\\
\email{\{tld,bulat\}@di.ku.dk}\\ \and Department of Odontology, University of Copenhagen
}
\maketitle              
%

%
%
%

\begin{abstract}
This paper describes our solution for the Dental Enumeration and Diagnosis on Panoramic X-rays Challenge at MICCAI 2023. Our approach consists of a multi-step framework tailored to the task of detecting and classifying abnormal teeth. The solution includes three sequential stages: dental instance detection, healthy instance filtering, and abnormal instance classification. In the first stage, we employed a Faster-RCNN model for detecting and identifying teeth.  In subsequent stages, we designed a model that merged the encoding pathway of a pretrained U-net, optimized for dental lesion detection, with the Vgg16 architecture. The resulting model was first used for filtering out healthy teeth. Then, any identified abnormal teeth were categorized, potentially falling into one or more of the following conditions: embedded, periapical lesion, caries, deep caries. The model performing dental instance detection achieved an AP score of 0.49. The model responsible for identifying healthy teeth attained an F1 score of 0.71. Meanwhile, the model trained for multi-label dental disease classification achieved an F1 score of 0.76. The code is available at https://github.com/tudordascalu/2d-teeth-detection-challenge.

\end{abstract}

\section{Introduction}
This article provides an overview of our solution submitted in the Dental Enumeration and Diagnosis on Panoramic X-rays Challenge held at MICCAI 2023~\cite{hamamci_dentex_2023}.

\section{Method}

\begin{figure}[t]
\includegraphics[width=\textwidth]{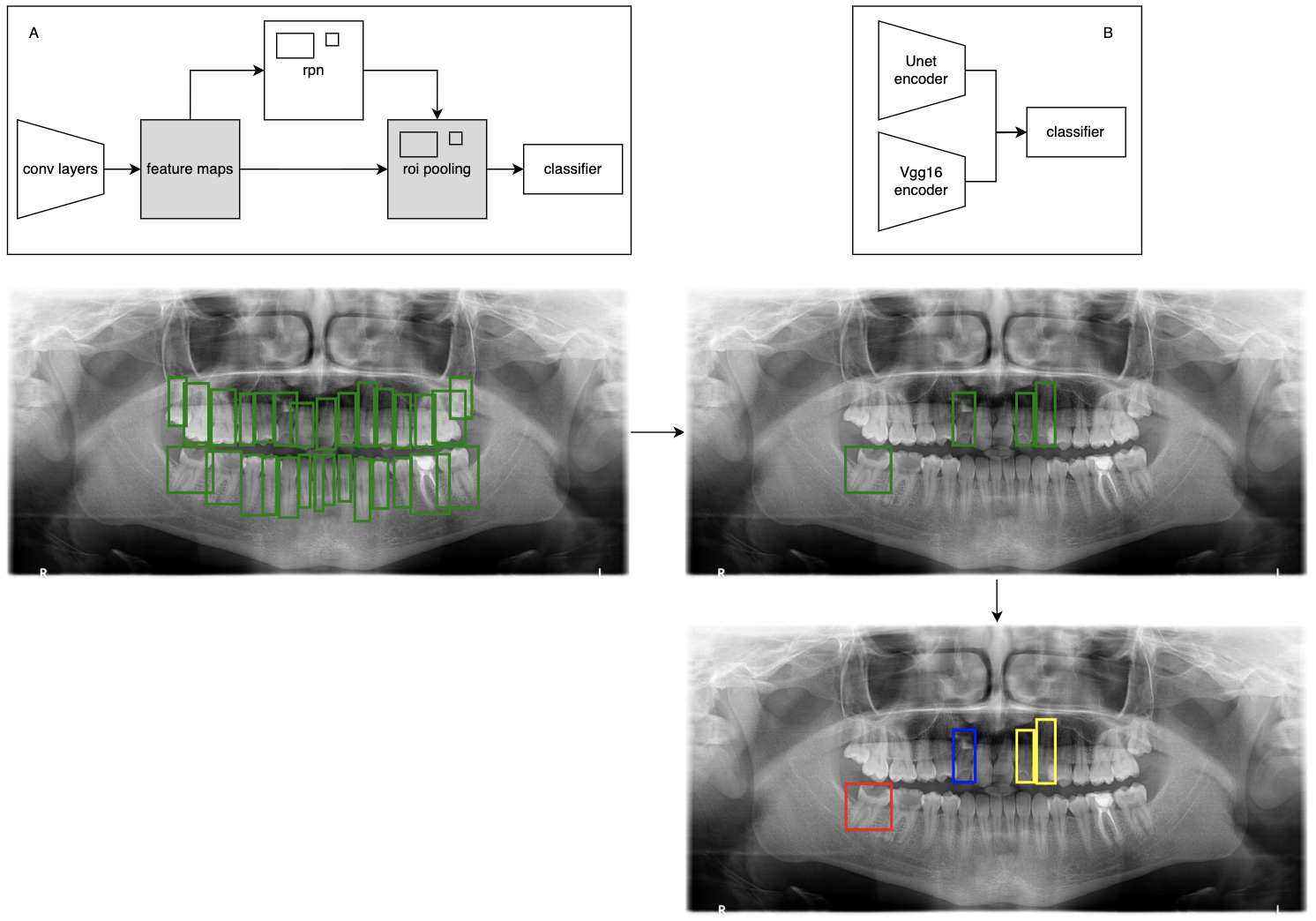}
\caption{The multi-step pipeline employed for detecting abnormal teeth. Initially, the Faster-RCNN (A) detects all teeth. Subsequently, a hybrid model incorporating features from both U-net and Vgg16 architectures (B) filters out healthy teeth. Finally, a model of the same architecture (B) classifies the abnormal teeth.} \label{fig_method}
\end{figure}

We proposed a multi-step framework designed for the detection and classification of abnormal teeth. This system consisted of three sequential stages: detection of dental instances, filtering of healthy instances, and classification of abnormal instances (Figure~\ref{fig_method}). 

In the initial phase, we employed a Faster-RCNN to identify dental instances from 2D panoramic X-rays, classifying them by quadrant and tooth number \cite{ren_faster_2016}. The model's outputs were bounding boxes, each paired with a confidence score and an encoded value representing the tooth and quadrant numbers.

In the following stages, we utilized a U-net model to segment both caries and periapical lesions from cropped tooth images \cite{ronneberger_u-net_2015}. The model was trained to minimize the mean of the Binary Cross Entropy loss and Dice loss. The encoding path of the trained U-net was integrated into the models used for (2) filtering out healthy instances and (3) distinguishing abnormal instances. These models presented a unified architecture, combining the U-net's encoding path with the Vgg16's feature extraction path~\cite{simonyan_very_2015}. Subsequently, the classification path of the Vgg16 classifier was adapted to handle the combined feature set sourced from both the U-net and Vgg16. The objective function used for training the models was the Binary Cross Entropy. The model designed to differentiate between healthy and abnormal teeth required cropped teeth images as input. It produced a binary label indicating the presence or absence of abnormalities. Teeth identified as abnormal were then processed by another model focused on their classification. This classification model was trained for multi-label categorization to accommodate teeth with multiple conditions.

\section{Experiment and results}
\subsection{Data}
The dataset used in the present work was introduced in the Dental Enumeration and Diagnosis on Panoramic X-rays Challenge. We leveraged a subset of 1039 Panoramic X-rays, with 634 X-rays associated with bonding boxes, teeth numbers, quadrant numbers corresponding to all the teeth, and 705 X-rays associated with bounding boxes, teeth numbers, quadrant numbers, disease type corresponding to abnormal teeth. The images were subjected to on-the-fly augmentation techniques such as horizontal flipping, brightness and contrast modifications, affine transformations, and random cutouts with dimensions reaching up to 80x80 pixels.

\subsection{Results}
In this section, the metrics presented include Average Precision at varying IoU thresholds for object detection and the F1 score for classification tasks. The results of our experiment are presented in Table~\ref{tab:results}. The Faster-RCNN model achieved an AP score of 0.49. The model differentiating healthy from unhealthy teeth obtained an F1 score of 0.71, while the classifier for abnormal teeth registered an F1 score of 0.76.

\begin{table}
\caption{The performance of the models performing (1) teeth detection and numbering, (2) healthy teeth classification, and (3) abnormal teeth classification.} \label{tab:results}
\begin{tabular}{|l|l|l|l|l|l|l|}
\hline
 Model & AP & AP75 & AP50 & F1 \\
 \hline
Teeth detection and numbering & 0.49 & 0.46 & 0.91 & - \\
Healthy teeth filtering & - & - & - & 0.71 \\
Abnormal teeth classification  & - & - & - & 0.76 \\
\hline

\end{tabular}
\end{table}

\bibliographystyle{splncs04}
\bibliography{mybibliography}

\begin{thebibliography}{1}
\providecommand{\url}[1]{\texttt{#1}}
\providecommand{\urlprefix}{URL }
\providecommand{\doi}[1]{https://doi.org/#1}

\bibitem{hamamci_dentex_2023}
Hamamci, I.E., Er, S., Simsar, E., Yuksel, A.E., Gultekin, S., Ozdemir, S.D.,
  Yang, K., Li, H.B., Pati, S., Stadlinger, B., Mehl, A., Gundogar, M., Menze,
  B.: {DENTEX}: {An} {Abnormal} {Tooth} {Detection} with {Dental} {Enumeration}
  and {Diagnosis} {Benchmark} for {Panoramic} {X}-rays (May 2023).
  \doi{10.48550/arXiv.2305.19112}, \url{http://arxiv.org/abs/2305.19112},
  arXiv:2305.19112 [cs]

\bibitem{ren_faster_2016}
Ren, S., He, K., Girshick, R., Sun, J.: Faster {R}-{CNN}: {Towards}
  {Real}-{Time} {Object} {Detection} with {Region} {Proposal} {Networks} (Jan
  2016), \url{http://arxiv.org/abs/1506.01497}, arXiv:1506.01497 [cs]

\bibitem{ronneberger_u-net_2015}
Ronneberger, O., Fischer, P., Brox, T.: U-{Net}: {Convolutional} {Networks} for
  {Biomedical} {Image} {Segmentation}. In: Navab, N., Hornegger, J., Wells,
  W.M., Frangi, A.F. (eds.) Medical {Image} {Computing} and
  {Computer}-{Assisted} {Intervention} – {MICCAI} 2015. pp. 234--241. Lecture
  {Notes} in {Computer} {Science}, Springer International Publishing, Cham
  (2015)

\bibitem{simonyan_very_2015}
Simonyan, K., Zisserman, A.: Very {Deep} {Convolutional} {Networks} for
  {Large}-{Scale} {Image} {Recognition}. arXiv:1409.1556 [cs]  (Apr 2015),
  \url{http://arxiv.org/abs/1409.1556}, arXiv: 1409.1556

\end{thebibliography}

\end{document}